%Paper: hep-ph/9208227
%From: HSU@huhepl.harvard.edu
%Date: Fri, 14 Aug 1992 17:16:48 -0400 (EDT)

%
%     Note to the reader:
%     body of paper is followed by postscript file containing figure,
%      which was generated using the "Draw" program on the NeXT.
%

\input harvmac
\pretolerance=750
\def\etal{{\it et al.}}

\def\prd#1#2#3{, {\sl Phys. Rev.} {\bf D#1} (19#2) #3}
\def\plb#1#2#3{, {\sl Phys. Lett.} {\bf #1B} (19#2) #3}
\def\npb#1#2#3{, {\sl Nucl. Phys.} {\bf #1B} (19#2) #3}

\Title{HUTP-92/A026} {Two Stage Phase Transition in Two Higgs Models}
\centerline{David Land\footnote{$^*$}
{This research was supported by the National Science Foundation under Grant
\#\nobreak PHY-8714654 and by the Texas National Research Laboratory
Commission under Grant \#\nobreak RGFY9106}}
\medskip
\centerline{Eric D. Carlson$^*$}
\bigskip\centerline{Lyman Laboratory of Physics}
\centerline{Harvard University}\centerline{Cambridge, MA 02138}

\vskip .3in
We present a mechanism in which models with two Higgs fields can undergo a two
stage phase transition as the temperature falls. The first stage is a
conventional second order (or weakly first order) transition in which the
symmetry is broken. Shortly thereafter follows a first order transition with
barrier penetration, bubble production and loss of thermal equilibrium. For
Higgs
potentials with CP violation, we show that the second stage of the transition
has all the features required for weak scale baryogenesis.

\Date{5/92}
%\draft

%\baselineskip=20pt plus 2pt minus 2pt

\newsec{Introduction}
The ${\rm SU(2)}_L \times {\rm U(1)_Y}$ symmetry of the standard model is
broken in the early universe as the temperature falls below the weak scale.
The order parameter for the phase transition is the real component of the
neutral Higgs field, which acquires a non-zero vacuum expectation value below
the transition temperature. The precise details of the transition are still
the subject of controversy, though the consensus seems to be that it is
weakly first order for Higgs masses greater than the experimental lower
bound \nref\linde{A. Linde, {\it Particle Physics and Inflationary Cosmology}
(Harwood Academic Publishers 1990) .}\nref\std{G.W. Anderson and L.J.
Hall\prd{45}{92}{2685} \semi M. Dine \etal \plb{283}{92}{319} \semi M.E.
Carrington \prd{45}{92}{2933}.}\nref\hsu{C.G. Boyd, D.E. Brahm,
S.D.H. Hsu,  {\tt   Work in Progress}.}\refs{\linde -\hsu}. That
is, at the transition, the Higgs vacuum expectation value (VEV) makes a jump
from zero to some very small fraction of its eventual zero temperature value.

Two Higgs models have also been investigated recently \nref\ruskies{A.I.
Bocharev, S.V. Kuzmin and M.E. Shaposnikov\plb{244}{90}{275}
\prd{43}{91}{369}.}\nref\others{N. Turok and J. Zadrozny\npb{358}{91}{471}
\semi B. Kastening, R.D. Peccei and X. Zhang\plb{266}{91}{413} \semi L.
McLerran \etal\plb{256}{91}{451} .}\refs{\ruskies, \others}, though the
whole range of parameter space has not been fully explored. Previous work has
concentrated on first finding the zero temperature absolute minimum of the
effective potential and then, in much the same way as has been done for the
single Higgs model, studying the evolution of this one minimum from high to
low temperatures. For certain combinations of parameters, one loop effective
potential calculations suggest that the transition can be first order
\ref\first{N. Turok and J. Zadrozny\npb{369}{92}{729}.}

In this paper, we make use of the large number of parameters in a two Higgs
model, and construct a potential in which the absolute minimum just after the
phase transition bears no relation to the absolute minimum at zero
temperature. At the transition point, the VEV is lured the ``wrong'' way, into
a  minimum which is only an absolute minimum for a short time close to the
critical temperature.  As the temperature continues to fall, it soon becomes a
false vacuum, because a second minimum begins to appear whose depth
quickly overtakes that of the first. At this point a tunneling process
will occur with production of bubble walls and loss of thermal equilibrium,
thus producing the main characteristics of a first order transition. As we
shall see, it is not difficult to ensure that the discontinuity in the VEV is
rather large. In \fig\nextdiag{Temperature development of the positions in
field space of the minima of the effective potential and the vacuum
expectation values of the scalars.}, these various stages of the transition
are indicated.

The electroweak phase transition, strictly speaking, is second order (or at
best weakly first order), and the first order phase transition occurs somewhat
later.  However, the jump in the VEV during the second stage can be large and
thus this mechanism could be useful in generating baryons at the electroweak
scale according to the scenarios recently proposed by Cohen, Kaplan and Nelson
in \nref\baryo{A.G. Cohen, D.B. Kaplan and A.E.
Nelson\plb{263}{91}{86}.}\nref\more{A.E. Nelson, D.B. Kaplan, A.G.
Cohen\npb{373}{92}{453}.}\refs{\baryo,\more}. They suggest that top quarks,
reflected from a bubble wall in a CP violating manner, produce a
hypercharge asymmetry which is converted into a baryon asymmetry in regions of
false vacuum where baryon violation is rapid. This asymmetry is swept into
the true vacuum when the bubble wall passes and, in order that it be
preserved, sphaleron processes in this region must be highly suppressed by a
large scalar VEV.

In section 2, we present the simplest example of the two stage transition in a
theory with just two scalar fields. In section 3, we apply the same ideas to
the more complicated example of the standard model with two Higgs doublets.
In section 4, we show that CP violation can be introduced into the potential
whilst retaining the other important features and thus that it is suitable for
weak scale baryogenesis.

\newsec{The Simplest Example}
Let us consider a theory of two real scalar fields $\phi_1$ and $\phi_2$ with
two discrete symmetries $\phi_1 \rightarrow -\phi_1$ and $\phi_2 \rightarrow -
\phi_2$.  We will think of $\phi_1$ as the standard Higgs, and $\phi_2$ as an
additional Higgs.  The Lagrangian is constrained to the form:
\eqn\scalarl{ {\cal L} = \half \partial^\mu \phi_1 \partial_\mu \phi_1 +
\half \partial^\mu \phi_2 \partial_\mu \phi_2 - V(\phi_1,\phi_2)\; ,}
where
\eqn\potl{ V(\phi_1,\phi_2) = k_1 \phi_1^4 + k_2 \phi_2^4 + k_3 \phi_1^2
\phi_2^2 - 2 \mu_1^2 \phi_1^2 - 2 \mu_2^2 \phi_2^2 \; .}
At zero temperature we will arrange the VEV to lie purely in the $\phi_1$
direction and to have a given value, in which case $\phi_1^2 = \mu_1^2 / k_1 =
(246 \, \rm GeV)^2$ for instance, and the value of the potential here is $V_1
= -\mu_1^4 / k_1$.
We can also arrange for there to be a second minimum, purely in the $\phi_2$
direction at $\phi_2 = \mu_2^2 / k_2$ with $V_2 = -\mu_2^4 / k_2$, which, if
it is not to be the absolute
minimum, must satisfy $|V_2| < |V_1|$. From the second derivatives of the
potential at the absolute minimum  it is easy to read off the zero temperature
mass spectrum : $m_1^2 = 8 \mu_1^2$; $m_2^2 = 2 (k_3/k_1)\mu_1^2  - 4
\mu_2^2$. Anticipating the generalization to the real world, we suppose that
there are some ``experimental'' constraints on these masses, let's say, that
$m_1, m_2 > 50 \,$GeV.

Next, we wish to make sure that the minimum which appears first during the
transition is the ``wrong'' one, i.e., the one in the $\phi_2$ direction. To
do this we must recompute the positions of the minima, this time as a function
of temperature. To see the effect that we have in mind, it will not be
necessary to go beyond leading order in high temperature corrections and for
this reason we use a simple method, put forward by Linde \linde, to calculate
the locations of the
minima. We compute thermally averaged equations of motion and assume that the
particle masses are sufficiently light in comparison to the temperature $T$
that we can use the result $\left< \phi^2 \right> = T^2/12$ as a good estimate
of the fluctuations in the scalar fields. In this way we obtain the following
pair of equations:
\eqn\min{\eqalign{& \left< \partial V \over \partial \phi_1 \right> = 0 = 4
k_1 \phi_1^3
+ \phi_1 \left[\alpha_1 T^2 + 2 k_3 \phi_2^2 - 4 \mu_1^2 \right] \; ; \cr &
\left< \partial V \over \partial \phi_2 \right> = 0 = 4 k_2 \phi_2^3
+ \phi_2 \left[\alpha_2 T^2  + 2 k_3 \phi_1^2 - 4 \mu_2^2 \right] \; ; \cr }}
where $\alpha_1 = k_1 + k_3 / 6$ and $\alpha_2 = k_2 + k_3 / 6$.
The non-zero $\phi_1$ and $\phi_2$ minima appear when the temperature drops
below
\eqn\tc{ {T_{c_1}}^2 = {4 {\mu_1^2} \over \alpha_1} \qquad \hbox{and} \qquad
 {T_{c_2}}^2 = {4 {\mu_2^2} \over \alpha_2} \; , }
respectively. In order that the VEV occur first in the $\phi_2$ direction, we
must make sure that $T_{c_1} < T_{c_2}$, even though we will end up arranging
them to be not too different.  Equations \min\ can have a third solution in
which both $\phi_1$ and $\phi_2$ are non-zero. In the case of interest to us,
however, this will be a saddle point rather than a minimum of the effective
potential.

Below these critical temperatures in this first stage of the transition, then,
the minima in the effective potential lie at:
\eqn\posa{ \phi_1^2 = {{4 \mu_1^2 - \alpha_1 T^2} \over {4 k_1} } \qquad ,
\qquad \phi_2 = 0 \; ;}
and
\eqn\posb{ \phi_1 = 0 \qquad , \qquad \phi_2^2 = {{4 \mu_2^2 - \alpha_2
T^2} \over {4 k_2} }  \; .}
As they develop, at some temperature $T_t$, the depths of the two minima
become equal and, below this temperature, \posa\ rather than \posb\ gives the
location of the true vacuum.  We assume here that bubble nucleation and
conversion to the low-temperature vacuum region occurs shortly thereafter.
$T_t$, thus, signals the onset of false vacuum decay and the second stage of
the transition. The tunneling temperature can be
calculated by substituting equations \posa\ and \posb\ into the temperature
corrected version of \potl\ which is:
\eqn\epotl{ V(\phi_1, \phi_2, T) = V(\phi_1, \phi_2) + T^2/2 \left[ \alpha_1
\phi_1^2 + \alpha_2 \phi_2^2 \right] \; .}
The values of the potential at the two minima are always given by $V_1 = - k_1
\phi_1^4$ and $V_2 = - k_2 \phi_2^4$ respectively, and so, if we want a large
jump in the size of the VEV during the jump, we should pick the ratio
$k_2 / k_1$ to
be large, since, when $V_1 = V_2$, the ratio of the VEV's goes as the fourth
root of this quantity.  This ratio is restricted, however, by the requirement
that the mass of the physical Higgs (coming from $k_1$) be not too small, and
the desire to avoid extremely non-perturbative couplings.  The tunneling
temperature is given by:
\eqn\tunt{ T_t^2 = {{4 (k_2^{1/2} \mu_1^2 - k_1^{1/2} \mu_2^2)} \over
{(k_2^{1/2} \alpha_1 - k_1^{1/2} \alpha_2)}} \; .}

It is not hard to find $k$'s and $\mu$'s which satisfy the three requirements
of large $k_2/k_1$, $T_{c_1} < T_{c_2}$, and $|V_1| > |V_2|$ at zero
temperature.  For instance, suppose we choose $k_1 = 0.01$, then $\mu_1 =
25\,$GeV to give a Higgs mass of $m_1 = 70\,$GeV.  To maximize $k_2/k_1$, we
choose $k_2$ at the edge of the perturbative regime, say $k_2 = 0.8$.  A large
value for $k_3$, such as $k_3=0.6$, will help us satisfy the temperature
constraint.  The value of $\mu_2$ is then constrained by the temperature and
true minimum constraints; a value of $\mu_2 = 72\,$GeV will suffice.  The
critical temperatures work out to $T_{c_1} = 151\,$GeV and $T_{c_2} =
152\,$GeV, and the mass of the second scalar is $m_2 = 233\,$GeV.  The
jump takes place at $T_t = 139\,$GeV, at which point the VEV changes
discontinuously from $32\,$GeV in the $\phi_2$ direction to $97\,$GeV in the
$\phi_1$ direction.

\newsec{The Two Higgs Doublet Standard Model}
We would now like to consider a more realistic possibility, where we include
Higgs doublets $\Phi_1$, $\Phi_2$ instead of singlets, and we will include
fermion couplings as well.  The symmetries $\phi_i \rightarrow - \phi_i$ will
be promoted to $\Phi_i \rightarrow - \Phi_i$, and we will allow such terms to
be softly broken.  We furthermore assume that only one of these two doublets
(specifically $\Phi_1$) is responsible for fermion masses, or at least, for
the top quark mass; such an assumption is usually necessary to avoid flavor
changing neutral currents.  The potential with these constraints is
conventionally written
\nref\higgs{J.F. Gunion, H.E. Haber, G. Kane and S. Dawson, {\it The Higgs
Hunter's Guide} (Addison-Wesley Publishing Company, 1990) .} \refs{\baryo -
\higgs}:
\eqn\oldpot{\eqalign{ V(\Phi_1, \Phi_2) \,\,\,\,\,\,\,\, = &\,\,\,\,\,\,\,
\lambda_1(\Phi_1^\dagger \Phi_1 - v_1^2)^2
+ \lambda_2(\Phi_2^\dagger \Phi_2 - v_2^2)^2 \cr & + \lambda_3[(\Phi_1^\dagger
\Phi_1 - v_1^2) + (\Phi_2^\dagger \Phi_2 - v_2^2)]^2 \cr & +
\lambda_4[(\Phi_1^\dagger \Phi_1)  (\Phi_2^\dagger \Phi_2) - (\Phi_1^\dagger
\Phi_2)  (\Phi_2^\dagger \Phi_1)] \cr & +  \lambda_5({\rm Re}(\Phi_1^\dagger
\Phi_2) - v_1 v_2 \cos\xi)^2 \cr & + \lambda_6({\rm Im}(\Phi_1^\dagger \Phi_2)
- v_1 v_2 \sin\xi)^2 \; . \cr}}
In this form, one can easily be blind toward the existence of minima other
than $\Phi_1^0 = v_1\,$; $\, \Phi_2^0 = v_2 e^{i \xi}$ (because some of the
coefficients above may be negative) and, for this reason, we find it more
convenient to rewrite the potential as:
\eqn\newpot{\eqalign{ V(\Phi_1, \Phi_2) \qquad = \qquad & 4 \{
k_1 (\Phi_1^\dagger \Phi_1)^2 + k_2 (\Phi_2^\dagger \Phi_2)^2 + k_3
(\Phi_1^\dagger \Phi_1) (\Phi_2^\dagger \Phi_2) \cr
& + k_4 [(\Phi_1^\dagger \Phi_1) (\Phi_2^\dagger \Phi_2) - (\Phi_1^\dagger
\Phi_2) (\Phi_2^\dagger \Phi_1)] \cr
& - \textstyle{1\over2} [k_5 (\Phi_1^\dagger \Phi_2)^2 + k_5^*(\Phi_2^\dagger
\Phi_1)]^2 \cr
& - \mu_1^2 \Phi_1^\dagger \Phi_1 - \mu_2^2 \Phi_2^\dagger \Phi_2 -
\mu_3^2 {\rm Re}(\Phi_1^\dagger \Phi_2) - \mu_4^2 {\rm Im}(\Phi_1^\dagger
\Phi_2) \} \;
. \cr}}
The presence of CP violation is signalled by the presence of $\mu_4^2$ and the
phase of $k_5$.  We can redefine the phase of one of the scalar fields to
eliminate, for example, the phase of $k_5$; we will assume from now on that
$k_5$ is real and positive (a similar phase choice was used to eliminate one
term in \oldpot).  The factor of $4$ is present so that when we make the
substitution:
\eqn\cmpnts{ \Phi_1 = {1 \over \sqrt2} \pmatrix{\chi_1 + i \eta_1 \cr
\phi_1 + i \psi_1 \cr} \qquad ; \qquad \Phi_2 = {1 \over \sqrt2}
\pmatrix{\chi_2 + i \eta_2 \cr \phi_2 + i \psi_2 \cr} \; ;}
the potential closely resembles that of the previous section.

The main difference from the model of the previous section is the much larger
number of degrees of freedom to be thermally averaged. We can thus expect the
coefficients $\alpha_i$ to be larger than before. Further, the Higgs
fields are not only self coupled but also coupled to the gauge bosons and the
fermions, though, of the latter, only the top quark has an appreciable Yukawa
coupling $\lambda_t$, and this, only to $\Phi_1$.

We will adjust the phases of the Higgs doublets so that the only fields in
\cmpnts\ which get a VEV are $\phi_1$, $\phi_2$ and $\psi_2$. Proceeding in
the same fashion as before, we can compute three thermally averaged equations
of motion:
\eqn\minima{\eqalign{ & \left< \partial V \over \partial \phi_2 \right> = 0 =
4 k_2 \phi_2^3 + \phi_2 \left[\alpha_2 T^2 + 2( k_3 - k_5) \phi_1^2 + 4 k_2
\psi_2^2 - 4 \mu_2^2 \right] - 2 \mu_3^2 \phi_1 \; ; \cr
& \left< \partial V \over \partial \psi_2 \right> = 0 = 4 k_2 \psi_2^3 +
\psi_2 \left[\alpha_2 T^2 + 2(k_3 + k_5) \phi_1^2 + 4 k_2 \phi_2^2 - 4 \mu_2^2
\right] - 2 \mu_4^2 \phi_1 \; ; \cr
&  \left< \partial V \over \partial \phi_1 \right> = 0 = 4 k_1 \phi_1^3 +
\phi_1
\left[\alpha_1 T^2 + 2( k_3 - k_5) \phi_2^2 + 2( k_3 + k_5) \psi_2^2 - 4
\mu_1^2
\right] \cr & \hskip 1.3in - 2 \mu_3^2 \phi_2 - 2 \mu_4^2 \psi_2 \; ; \cr }}
where
\eqn\consts{\eqalign { & \alpha_1 = 2 k_1 + {\textstyle{2\over3}} k_3  +
{\textstyle{1\over3}}k_4 + {e^2 (1 + 2 \cos^2\theta_W) \over \sin^2 2
\theta_W}
+ 2 \lambda_t^2 \; ; \cr
& \alpha_2 = 2 k_2 + {\textstyle{2\over 3}} k_3 + {\textstyle{1\over 3}}k_4 +
{e^2 (1 + 2 \cos^2\theta_W) \over \sin^2 2 \theta_W} \; , \cr} }
where $e$ and $\theta_W$ are the electromagnetic coupling and electroweak
mixing angle respectively.
For the moment, let us restrict our attention to the case when
$k_4 = k_5 = 0$ and $\mu_3 = \mu_4 = 0$ so that there is no CP violation in
the potential and the VEVs may both be taken to be real. With this
simplification the equations \minima\ become the same as \min\ except that, as
expected, there are more contributions to $\alpha_i$. We will take $\lambda_t
= 0.5$ corresponding to a zero temperature top mass of $123\,$GeV.

The analysis of the transition is now almost identical to that described in
the previous section. The eventual VEV is still $\phi_1 = 246\,$GeV, so if we
choose as before $k_1 = 0.01$, $\mu_1 = 25\,$GeV, then again $m_1 = 70\,$GeV.
Once more, choosing $k_2 = 0.8$, and this time $k_3 =
0.2$, we then immediately obtain from \consts\ $\alpha_1 = 0.987$ and
$\alpha_2 = 2.07$. From \tc\ we then obtain $T_{c_1} = 50\,$GeV. The choice
$\mu_2 = 43\,$GeV will then satisfy the remaining constraints. We find $m_2 =
130\,$GeV
to be the mass of the other neutral scalar as well as the charged scalars and
the pseudoscalar. They are all equal in mass because of the residual symmetry
in the Higgs sector when no component of $\Phi_2$ has a VEV. From \tc\ we
find that the first transition occurs at $T_{c_2} = 60\,$GeV and from \tunt\
we see that this is followed by a tunneling transition at $T_t =
46\,$GeV, at
which point, the VEV changes from $30\,$GeV in the $\phi_2$ direction to
$90\,$GeV in the $\phi_1$ direction, as we can see by substitution in \posa\
and \posb. Some of the details of this transition are displayed in \nextdiag.

Reintroduction of certain of the missing terms causes no substantial changes.
The main effect of a positive $k_4$ is to increase the charged scalar masses;
these do not
interest us at the moment, so we will leave this parameter vanishing.  If we
increase $k_5$ to a non-zero value, the equations \minima\ change, but not
their solution.  The extra scalar and pseudo-scalar masses will also be split,
but these are so large that we can allow $k_5$ to be almost as great as $k_3$
without running into
any difficulties with experiment.  The main effect of $k_5$ is to assure that
the phases of $\Phi_1$ and $\Phi_2$ are aligned (that is, $\psi_2 =0$) when
they are both non-zero, so, even with these complications, we expect the
phases of the two Higgs fields to be aligned in the transition region
comprising the middle of the bubble wall.
This term, however, will have very little effect on the regions near either of
the two local minima, when one of the fields vanishes, and it is these regions
that constitute the top and bottom of the wall. The effects of the  $\mu_3^2$
and $\mu_4^2$ terms will be studied in the next section.

\newsec{CP Violation and Weak Scale Baryogenesis}
We wish, now, to show that the transition which we have described may be
useful in generating a baryon excess at the weak scale.  The potential already
has the necessary out of equilibrium phase transition, and electroweak
interactions can violate baryon number, but we have not introduced CP
violation.To achieve this, we will now reinstate $\mu_3$ and $\mu_4$;
however, we will keep them
sufficiently small (say ${\cal O}(1\, \rm GeV)$) that we only need consider
perturbations about the
positions of the minima computed in section 3. We must return to \minima\ and
calculate the corrected positions of the minima. Let's see what happens to the
minimum in the $\phi_1$ direction. It will pick up small
components in the $\phi_2$ and $\psi_2$ directions. The first two equations in
\minima\ may be combined for non zero $\phi_1$ to yield:
\eqn\phase{2 k_5 \phi_1 \phi_2 \psi_2 - \mu_4^2 \phi_2 + \mu_3^2 \psi_2 =
0\; .}
The first term here is second order in the small fields and so we obtain for
the tangent of the CP violating phase: $\psi_2 / \phi_2 \approx \mu_4^2 /
\mu_3^2$. Similarly, the other minimum will pick up a small component in the
$\phi_1$ direction. Either of the first two equations in \minima\ can be used
to determine the magnitude of $\Phi_2^0$ by ignoring the small $\phi_1$ terms.
With non-zero $\phi_1$, however, we are no longer free to choose $\psi_2$ to
vanish, and hence the CP
violating phase to be zero, and instead must determine it from equation
\phase. Once again, the first term is negligible because $\phi_1$ is
perturbatively small and we obtain as before: $\psi_2 / \phi_2 \approx \mu_4^2
/ \mu_3^2$.
The small component $\phi_1$ is then given approximately by the solution of
the cubic derived from the third equation in \minima,
where, for $\phi_2$ and $\psi_2$ we use the values we have just calculated
from the first pair of equations in \minima.

The phase structure of a bubble wall can thus be estimated when we
recall from the previous section that in the middle of the wall, where both
$\phi_1$ and $\phi_2$ are large, the phase tends to be driven to zero so as
to minimise the $k_5$ term in the potential. We can thus typically expect the
variations in phase experienced by top quarks during reflection from the wall
to be $\Delta \theta = {\cal O} (\arctan \mu_4^2 / \mu_3^2)$.  Naively, in the
spirit of \baryo, we might expect the preferential reflection of hypercharge
to be proportional to $\Delta \theta$\footnote{$^1$}{We should note that, in
the limit of the CP violating parameters tending to zero, the preferential
reflection of top quarks with definite hypercharge will not be proportional to
$\Delta \theta$.  This is a reason for wishing to obtain a description of the
transition without making these parameters too small.}, which may lead,
ultimately, to a baryon asymmetry.  Thus we see that the introduction of small
$\mu_3$ and $\mu_4$ values may result in net baryon production at the phase
transition wall.

For somewhat larger values of $\mu_3$ and $\mu_4$, we do not expect the
essential features of the transition to be altered, but to obtain accurate
results for the quantities involved, it becomes necessary to find the relevant
solutions of \minima\ by numerical methods.

Recall from section 3, that at the jumping temperature, we
have the following ratios of VEV to temperature : $v/T_t = 0.67$ in the false
vacuum; $v/T_t = 2.0$ in the true vacuum. Baryon violation is suppressed in
its rate by a factor of $\exp( - {4 \pi \over g_W} v/T)$ \ref\sphaleron{N.S.
Manton\prd{21}{80}{1591} \semi N.S. Manton\prd{28}{83}{2019} \semi F.R.
Klinkhamer and N.S. Manton\prd{30}{84}{2212} \semi V.A. Kuzmin, V.A. Rubakov
and M.E. Shaposhnikov\plb{155}{85}{36} \semi P. Arnold and L.
McLerran\prd{36}{87}{581}\prd{37}{88}{1020} .}, where $g_W$ is the ${\rm
SU(2)}_L$ gauge coupling. When the VEV makes its
jump during the second stage of the
transition, the rate is thus reduced by a factor of approximately $4 \times
10^{-12}$, to a value which is slow enough to ensure that any baryon asymmetry
created will not be subsequently dissipated. In false vacuum regions, in
contrast, the baryon violation, whilst not completely unsuppressed, can still
be rapid enough to convert hypercharge to baryon number before the bubble wall
sweeps past. To generate the observed baryon asymmetry, therefore, we favour
bubble walls with low velocities and a large CP violating parameter $\Delta
\theta = {\cal O} (\arctan \mu_4^2 / \mu_3^2)$.

\newsec{Conclusion}
In exploring the parameter space of two Higgs models, we have found that the
symmetry breaking process may occur in two steps, the second of which produces
many of the features of a strongly first order phase transition. We have shown
that this may be important in certain weak scale baryogenesis scenarios, where
it is crucial that the baryons do not disappear immediately subsequent to the
transition, as happens if it is only weakly first order.

Calculation of higher order thermal loop graphs to obtain an improved
effective potential will only yield small corrections since the interesting
stage of the transition occurs well away from the origin in field space. It
may be interesting, however, to work out more accurately, just how large a CP
violating phase can be obtained by adjusting the various parameters.
\vfill\eject

\listrefs
\listfigs

\bye